\documentstyle[twocolumn]{article}
\makeatletter
\def\@normalsize{\@setsize\normalsize{10pt}\xpt\@xpt
\abovedisplayskip 10pt plus2pt minus5pt\belowdisplayskip 
\abovedisplayskip \abovedisplayshortskip \z@ 
plus3pt\belowdisplayshortskip 6pt plus3pt 
minus3pt\let\@listi\@listI}
\def\subsize{\@setsize\subsize{12pt}\xipt\@xipt}
\def\section{\@startsection {section}{1}{\z@}{1.0ex plus 
1ex minus .2ex}{.2ex plus .2ex}{\large\bf}}
\def\subsection{\@startsection {subsection}{2}{\z@}{.2ex 
plus 1ex} {.2ex plus .2ex}{\subsize\bf}}
\makeatother
\begin{document}
\date{}
\title{\Large\bf Three Logistic Models for the Two-Species Interactions: \\ 
Symbiosis, Predator-Prey and Competition}
\author{\begin{tabular}[t]{c@{\extracolsep{8em}}c} 
R. L\'opez-Ruiz$^\dag$ and D. Fournier-Prunaret$^\ddag$ \\
 \\
     \small   $^\dag$ DIIS-BIFI, Universidad de Zaragoza, \\
	 \small  Campus San Francisco, 50009 - Zaragoza (Spain). \\
	 \small	$^\ddag$ SYD-LESIA, INSA Toulouse, \\
     \small   Campus Rangueil, 31077 - Toulouse Cedex (France). \\
\end{tabular}}
\maketitle
\thispagestyle{empty}
\subsection*{\centering Abstract}
{\em
If one isolated species is supposed to evolve fo-llowing 
the logistic mapping, then we are tempted to 
think that the dynamics of two species can be expressed 
by a coupled system of two discrete logistic equations. 
As three basic relationships between two species are present 
in Nature, namely symbiosis, predator-prey and competition,
three different models are obtained. Each model is a cubic 
two-dimensional discrete logistic-type equation with its own dynamical 
properties: stationarity, periodicity, quasi-periodicity and chaos. 
Furthermore, these models could be consi-dered as the basic ingredients 
to construct more complex interactions in the ecological networks. 
}

\section{Introduction}

If $x_n$ represents the population of an isolated species 
after $n$ generations, let us suppose this variable 
is bounded in the range $0<x_n<1$. 
A simple model that gives account of its evolution is
the so-called logistic map \cite{may},      
\begin{equation}
x_{n+1}=\mu\; x_n(1-x_n),
\end{equation}
where $0<\mu<4$ in order to assure $0<x_n<1$.
The term $\mu x_n$ controls the {\it activation or expanding phase}, 
where $\mu$ expresses the {\it growth rate}. 
The term $(1-x_n)$ inhibits the overcrowding, then it is controlling
the {\it inhibition or contracting phase}. 
The continuous version of this model was 
originally introduced by Verhulst in the 
nineteenth century as a counterpart to the Malthusian 
theories of human overpopulation.   

\section{The Models}

If two species $(x_n,y_n)$ are now living together \cite{lopez},
each evolves following a logistic-type dynamics,
\begin{eqnarray}
x_{n+1} & = & \mu_x(y_n)\;x_n(1-x_n), \\
y_{n+1} & = & \mu_y(x_n)\;y_n(1-y_n). 
\end{eqnarray}
The interaction between species causes
the growth rate $\mu(z)$ to vary with time, then $\mu(z)$ depends on the 
po-pulation size of the others and on a positive constant $\lambda$ that 
measures the strength of the mutual interaction.  
The simplest choice for this growth rate can be 
a linear increasing $\mu_1$ or decreasing $\mu_2$ function expanding 
at the parameter interval where
the logistic map shows some activity, that is $\mu\in(1,4)$. Thus,
\begin{eqnarray}
\mu_1(z) & = & \lambda\;(3z+1), \\
\mu_2(z) & = & \lambda\;(-3z+4).
\end{eqnarray}
Then we have: \newline
(1) The {\it symbiosis} originates a symmetrical coupling due 
to the mutual benefit, then $\mu_x=\mu_y=\mu_1$. \newline
(2) The {\it predator-prey interaction} is based on the benefit/damage relationship established
between the predator and prey, respectively, then $\mu_x=\mu_1$ and $\mu_y=\mu_2$. \newline
(3) The {\it competition} between species causes the contrary symmetrical coupling, 
then $\mu_x=\mu_y=\mu_2$.

\section{Summary and Conclusions}
Three discrete two-dimensional logistic systems are proposed
to model the basic interactions between pairs of species.
These could be considered for future studies as the bricks necessary to built more
complex interaction networks among species. 


\end{document}